\documentclass[useAMS,usenatbib]{mn2e}
\usepackage{exscale}
\usepackage{graphicx}
\usepackage{natbib}
\usepackage{rotating}
\usepackage{rotate}
\usepackage{afterpage}
\usepackage{fancyhdr}
\usepackage{amsmath}
\usepackage{amssymb}
\usepackage{relsize}
\usepackage{lscape}
\usepackage{epsfig}
\usepackage{sidecap}
\usepackage{tabularx}
\usepackage{longtable}
\usepackage{ltxtable}
\usepackage{txfonts}
\usepackage{natbib}
\usepackage{bm}

\def\apj{ApJ}%
\def\apjl{ApJ}%
\def\apjs{ApJS}%
\def\aap{A\&A}%
\def\mnras{MNRAS}%
\def\prd{Phys.~Rev.~D}%
\def\nat{Nature}%
\newcommand{\ti}{\textit}

\newcommand{\bea}{\begin{eqnarray}}
\newcommand{\be}{\begin{equation}}
\newcommand{\ben}{\begin{enumerate}}
\newcommand{\bi}{\begin{itemize}}
\newcommand{\eea}{\end{eqnarray}}
\newcommand{\ee}{\end{equation}}
\newcommand{\ei}{\end{itemize}}
\newcommand{\een}{\end{enumerate}}

\newcommand{\vpi}{\mathbf \pi}
\newcommand{\vecd}{\mathbf d}
\newcommand{\matC}{\mathbf C}
\newcommand{\matQ}{\mathbf Q}

\newcommand{\rrEB}{\mathcal R \mathcal R_\mr{E/B}}

\newcommand{\om}{\Omega_\mr m}
\newcommand{\omb}{\Omega_\mr b}
\newcommand{\sig}{\sigma_8}
\newcommand{\ns}{n_s}
\newcommand{\w}{w_0}

\newcommand{\pd}{P_{\delta}}
\newcommand{\pe}{P_\mr E}

\newcommand{\vt}{\vartheta}

\newcommand{\eps}{\epsilon}

\newcommand{\mr}{\mathrm}

\renewcommand{\d}{{\rm d}}

\def\tmin{{\vt_{\rm min}}}
\def\tmax{{\vt_{\rm max}}}
\def\tplog{T_{+n}^\mr{log}}
\def\tmlog{T_{-n}^\mr{log}}
\def\tpmlog{T_{\pm n}^\mr{log}}
\def\xipm{\xi_\pm}
\def\xip{\xi_+}
\def\xim{\xi_-}
\def\kmax{k_\mr{max}}

\title[Weak lensing statistics from the Coyote Universe]{Weak lensing statistics from the Coyote Universe}
\author[Tim Eifler]{Tim Eifler$^{1}$\thanks{E-mail:
teifler@mps.ohio-state.edu} \\
$^{1}$ Center for Cosmology and Astro-Particle Physics, The Ohio State University , 191 W. Woodruff Ave, Columbus, 43210 OH, USA}

\begin{document}
\date{accepted received}

\pagerange{\pageref{firstpage}--\pageref{lastpage}} \pubyear{2010}

\maketitle

\label{firstpage}

\begin{abstract}
{Analyzing future weak lensing data sets from KIDS, DES, LSST, Euclid, WFIRST requires precise predictions for the weak lensing measures. In this paper we present a weak lensing prediction code based on the Coyote Universe emulator. The Coyote Universe emulator predicts the (non-linear) power spectrum of density fluctuations ($P_\delta$) to high accuracy for $k \in [0.002;3.4] \, \mr{h/Mpc}$ within the redshift interval $z \in [0;1]$, outside this regime we extend $P_\delta$ using a modified Halofit code.\\
This pipeline is used to calculate various second-order cosmic shear statistics, e.g., shear power spectrum, shear-shear correlation function, ring statistics and COSEBIs (Complete Orthogonal Set of EB-mode Integrals), and we examine how the upper limit in $k$ (and $z$) to which $P_\delta$ is known, impacts on these statistics. For example, we find that $\kmax \sim 8$ $\mr{h/Mpc}$ causes a bias in the shear power spectrum at $\ell \sim 4000$ that is comparable to the statistical errors (intrinsic shape-noise and cosmic variance) of a DES-like survey, whereas for LSST-like errors $\kmax \sim 15$ $\mr{h/Mpc}$ is needed to limit the bias at $\ell \sim 4000$. \\
For the most recently developed second-order shear statistics, the COSEBIs, we find that 9 modes can be calculated accurately knowing $P_\delta$ to $\kmax=10$ $\mr{h/Mpc}$. The COSEBIs allow for an EB-mode decomposition using a shear-shear correlation function measured over a finite range, thereby avoiding any EB-mode mixing due to finite survey size. We  perform a detailed study in a 5-dimensional parameter space in order to examine whether all cosmological information is captured by these 9 modes with the result that already 7-8 modes are sufficient.
}
\end{abstract}


\section{Introduction}
Weak lensing by large-scale structure, so-called cosmic shear, was first detected in 2000 independently by several groups \citep{bre00,kwl00,wme00,wtk00}, and has recently progressed to an important tool in cosmology. Latest results \citep[e.g.][]{wmh05,smw06,hmv06,ses06,hss07,mrl07,fsh08,shj10} already indicate its great ability to constrain cosmological parameters which will be enhanced by large upcoming ground based surveys like KIDS\footnote{http://www.astro-wise.org/projects/KIDS/}, DES\footnote{www.darkenergysurvey.org/}, LSST\footnote{http://www.lsst.org/lsst}, and satellite missions Euclid\footnote{sci.esa.int/euclid/} and WFIRST\footnote{http://wfirst.gsfc.nasa.gov/}.\\
In order to meet the requirements for analyzing high precision data sets from these future surveys cosmic shear needs to overcome several challenges. On the observational side accurate shape measurements \citep[see][for latest developments]{hwb06,mhb07,bbb09,ber10} and photometric redshift information \citep{beh10,hac10} probably pose the most challenging problems. On the astrophysical side intrinsic alignments can mimic a shear signal \citep{his04,mhi06,jma10,mbb10} and need to be either removed \citep{kis03,jos08,jos09}, or modeled carefully in the subsequent analysis \citep{brk07,job09,scb10,kbs10}.\\
An important step to quantify and check for these systematics is the decomposition into E- and B-modes, where, to leading order, gravitational lensing only creates E-modes. In principle B-modes can arise from the limited validity of the Born approximation \citep{jsw00,hhw09} or redshift source clustering \citep{svm02}. Predictions coming from numerical simulations differ on the impact of these effects \citep[e.g.][]{hrh00,cnp01,jin02}, however, the observed B-mode amplitude is higher than one would expect from the foregoing explanations. Lensing bias \citep{srd09} is another possible source of B-modes; again the amplitude compared to the E-mode is small \citep{krh10}. Most likely, a B-mode detection indicates remaining systematics in the observation and data analysis, in particular an insufficient PSF-correction. \\
There exist several methods to perform the EB-mode decomposition; as a general classification they can be separated into real- and Fourier-space methods, where the latter are inspired by the pseudo-CL formalism invented to analyze the CMB polarization power spectrum \citep{hgn02,bct05}. A finite survey area and masking effects introduce a mixing of E- and B-modes in the CL's (also called \ti{leakage}), which prohibits a clean separation \citep{lct02}. The effect cannot be removed completely, however for CMB polarization the arising leakage B-mode can be suppressed \citep{lew03,smi06,kin10} to a level that enables for a possible detection of primordial B-modes (depending on the primordial tensor-to-scalar ratio).\\
\cite{hth10} extend the Pseudo-CL formalism to weak lensing, and test it on ray-tracing simulations, finding an B-mode leakage at the percent level and below depending on the Fourier mode $\ell$. However, the strength of this effect depends on masking and survey geometry and it needs to be examined on a case by case basis. Compared to CMB polarization additional difficulties arise in lensing when calculating the errors/covariances of the Pseudo CL's. For CMB polarization this covariance can be expressed in terms of the Pseudo CL's themselves assuming that the underlying field is Gaussian \citep{chc05}. This assumption is justified for the CMB field (if primordial non-Gaussianity is small/zero), however it is not true for the late time shear field where non-linear structure growth leads to non-negligible non-Gaussian effects \citep{whu00,svh07,taj09,sht09}. Here, higher order terms arise in the Pseudo CL shear covariance and the impact of EB-mode leakage on these terms is unknown.\\
Several real space EB-decomposing methods suffer from EB-mode mixing as well, e.g., the aperture mass dispersion, the shear E-mode correlation function, and the shear dispersion \citep{kse06}. These statistics can be calculated in terms of the shear two-point correlation function (2PCF), however all three measures need information on scales outside the interval $[\tmin;\tmax]$ of the measured 2PCF. This information can of course be modeled using a theoretical 2PCF, however this biases the results towards the cosmological model assumed in the 2PCF extension. The ring statistics \citep{sck07,esk10,fuk10} and more recently the COSEBIs \citep[][hereafter SEK10]{sek10} provide a new method to perform an EB-mode decomposition using a 2PCF measured over a finite angular range, thereby avoiding any EB-mode leakage/mixing.\\
Independent of whether cosmic shear data is analyzed in Fourier or real space both methods rely on accurate predictions for the corresponding shear measure in order to constrain cosmological models to the desired precision. \cite{hut05} find that one needs to know the power spectrum of density fluctuations ($P_\delta$) to $<$1\% accuracy over a range of 1-10 h/Mpc in order to obtain cosmic shear predictions that are sufficiently accurate for LSST. The Halofit fitting formula for $P_\delta$ described in \cite{smp03} is only accurate to $\sim$10\% (depending on cosmology and scale). The Coyote Universe emulator \citep{heh09,hww10,lhw10} has significantly improved on this issue; it models $P_\delta$ with percent accuracy in a 5 dimensional parameter space over the range $k \in [0.002;3.4] \, \mr{h/Mpc}$  within $z \in [0;1]$.\\
This paper has two goals: First, we built a predictions pipeline for second-order shear measures based on the Coyote Universe emulator and a modified Halofit. For improved weak lensing predictions from future numerical simulations we examine to which scale $k$ and redshift $z$ the density power spectrum must be known to model various second-order shear statistics to the desired accuracy for DES and LSST. The second goal is to examine the COSEBIs in a 5-dimensional parameter space. As outlined in SEK10 the COSEBIS do not only offer a check for B-mode systematics, moreover the authors argue that the COSEBIs' E-mode is the quantity that should enter a cosmic shear likelihood analysis. The 2PCF itself in particular should not be used for this purpose. The reason for this is that even when finding no B-modes in any EB-mode decomposition tests, the 2PCF can still be affected by B-modes that 1) mimic a constant or linear shear field 2) are localized in the power spectrum but present on scales ($\ell$) larger than the survey area.\\
The (logarithmic) COSEBIs are designed to contain all EB-decomposable cosmological information, and compress it into few data points (modes). SEK10 examined this in the two-dimensional parameter space $\om$ vs. $\sig$; we extend their analysis to 5 parameters which correspond to the cosmological parameter space covered by the Coyote Universe.\\
The paper starts with a short description of our weak lensing prediction pipeline; here and in Sect. \ref{sec:cosebis}, we also determine the requirements for future numerical simulations needed for sufficiently accurate weak lensing predictions for DES and LSST. In Sect. \ref{sec:like_analysis} we examine the performance of the COSEBIs in a 5-dimensional parameter space, in particular we address the questions 1) how many modes of the COSEBIs need to be included before the cosmological information saturates and 2) how does this saturation limit compare to the case of a pure E-mode 2PCF (which we can of course simulate but is never guaranteed in a real data set). We discuss our findings and conclude in Sect. \ref{sec:conclusions}.

\section{The second-order shear statistics prediction code}
\label{sec:predictions}
The code is based on the Coyote Universe power spectrum emulator \citep{lhw10}, which in its newest version emulates $\pd$ over the range $k \in [0.002;3.4]$ h/Mpc within $z \in [0;1]$. For the higher $k$ and $z$ range we calculate the linear power spectrum from an initial Harrison-Zeldovich power spectrum ($P_{\delta}(k) \propto k^{n_\mr s}$) using the transfer function of \cite{ebw92}. We account for the non-linear structure growth using the formula of \cite{smp03} and then scale the result such that the transition to the Coyote Universe power spectrum is smooth. More precisely, we derive the scaling factor by taking the ratio of $\pd$(Coyote)/$\pd$(Halofit) for the largest $k$ of the Coyote
emulator. For all $k>\kmax$(Coyote) we multiply the amplitude of Halofit with this factor. \\
The main difficulty when co-implementing the emulator and Halofit is that the former automatically calculates $H_0$ such that it fits the WMAP-5year constraints on the angular scale of the acoustic peak \citep{kdn09}, which implies that our code cannot vary $\om$, $\omb$ independently of $H_0$. 
\subsection{The shear power spectrum }
\label{sec:pe}
\begin{figure}
\includegraphics[width=9cm]{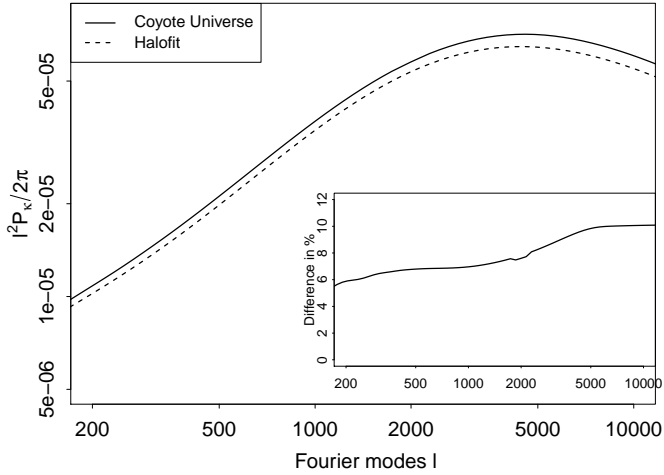}
  \caption{The shear power spectrum when calculated with Halofit or from the new Coyote Universe/modified Halofit code.}
         \label{fi:xi_pe_compare}
\end{figure}
The shear power spectrum $\pe$ can be expressed as a an integral over the density power spectrum $\pd$ through Limber's equation
\be
\label{eq:pdeltatopkappa}
\pe (\ell) = \frac{9H_0^4 \om^2}{4c^4} \int_0^{w_\mr h} 
\mr d w \, \frac{g^2(w)}{a^2(w)} \pd \left(\frac{\ell}{f_K(w)},w \right) \,,
\ee
where $\ell$ is the 2D wave vector perpendicular to the line of sight, $w$ denotes the comoving coordinate, $w_\mr h$ the comoving coordinate of the horizon, $a(w)$ is the scale factor, $f_K(w)$ the comoving angular diameter distance.\\
The weight factor $g$ is defined as an integral over the redshift distribution of source galaxies $n(w(z))$
\be
\label{eq:redshift_distri}
g(w) = \int_w^{w_{\mr h}} \mr d w' n (w') \frac{f_K (w'-w)}{f_K (w')} \,.
\ee
In the calculation of $P_\mr E$ we choose a redshift distribution of source galaxies similar to that for the CFHTLS described in \cite{bhs07}
\be 
\label{redshiftben}
n(z)=\frac{\beta}{z_0 \Gamma \left( \left(1+\alpha \right)/\beta \right)} \left( \frac{z}{z_0}\right)^\alpha \exp \left[ - \left(  \frac{z}{z_0} \right)^\beta \right]\,,
\ee 
with $\alpha=1.197$, $\beta=1.193$, $z_0=0.555$, and a cut-off at $z_\mr{max}=4$. \\
Figure \ref{fi:xi_pe_compare} compares the shear power spectrum calculated from our modified code compared to using Halofit only. The assumed cosmology corresponds to the best fit cosmology of the WMAP 7-year analysis \citep{ksd10}, i.e. $\om=0.272, \Omega_\Lambda=0.728, \sig=0.807$, $h=0.703$, $\Omega_\mr b=0.045$, $n_\mr s=0.961, w_0=-1$, and $w_a=0 $. These parameters also define our fiducial cosmology used for the fiducial data vector in the likelihood analysis (Sect. \ref{sec:like_analysis}). A direct comparison to power spectra from raytracing simulations is unfortunately not possible due to the fact that $\om$, $\Omega_\mr b$ cannot be chosen independently, hence there was no raytracing simulation with the ``correct'' cosmology at hand.\\

\subsection{Required accuracy in $\pd$ for future weak lensing surveys}
\begin{figure}
\includegraphics[width=9cm]{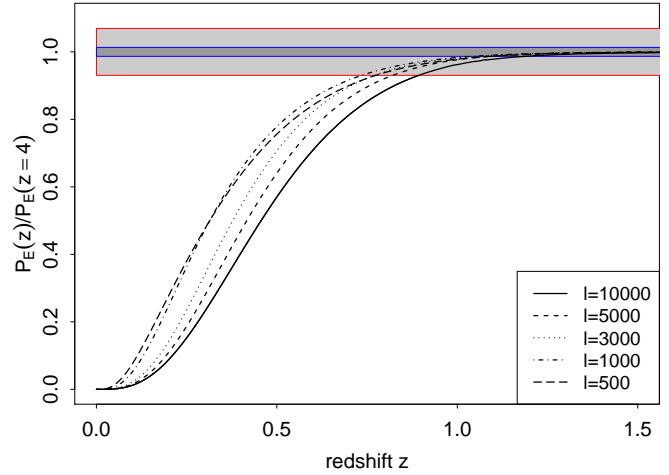}
\caption{The fraction of $\pe(z)/\pe(z=4)$ for various different $\ell$ where $\pe(z)$ is calculated from Limber's equation but with $w(z)$ being the upper limit of the integral. We see that the shear power spectrum can be calculated accurately when knowing $\pd$ only up to $z=1.5$. This will of course depend on the assume redshift distribution. The error bars are calculated for a DES-like survey (light shaded), and for a LSST-like survey (dark shaded) at $\ell=1000$. }
\label{fi:integrand_pe}
\end{figure}

\begin{figure}
\includegraphics[width=8cm]{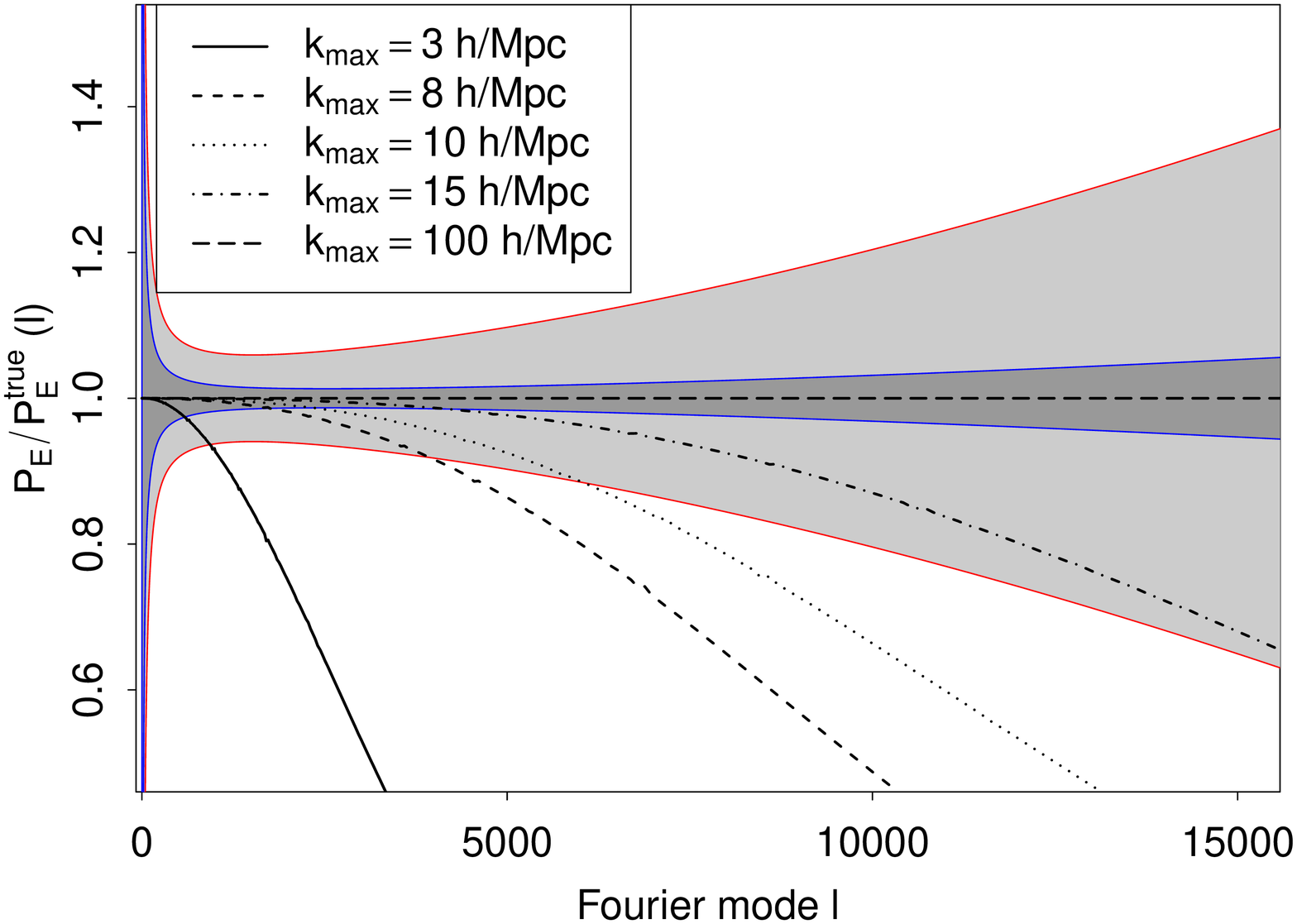}
\includegraphics[width=8cm]{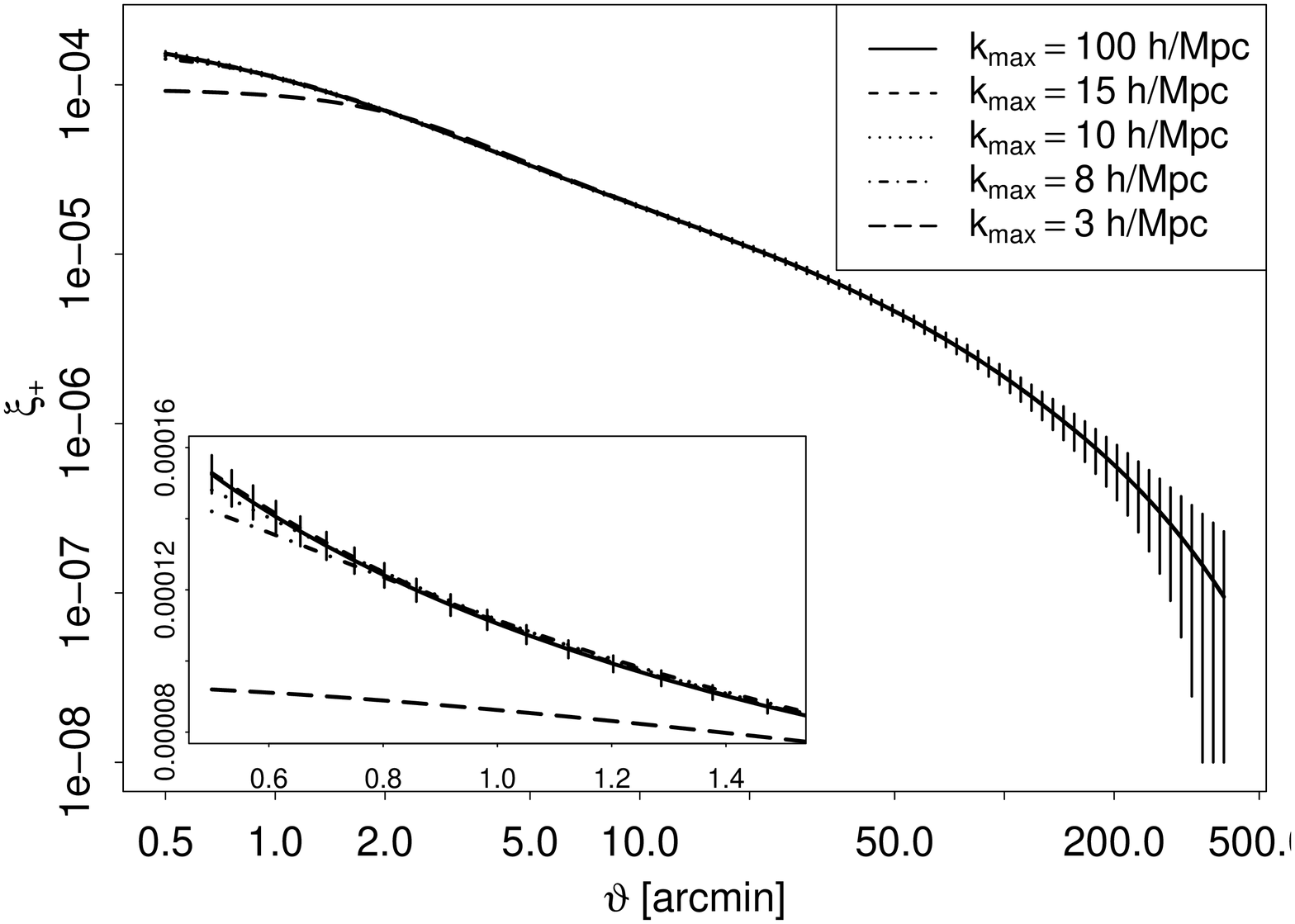}
\includegraphics[width=8cm]{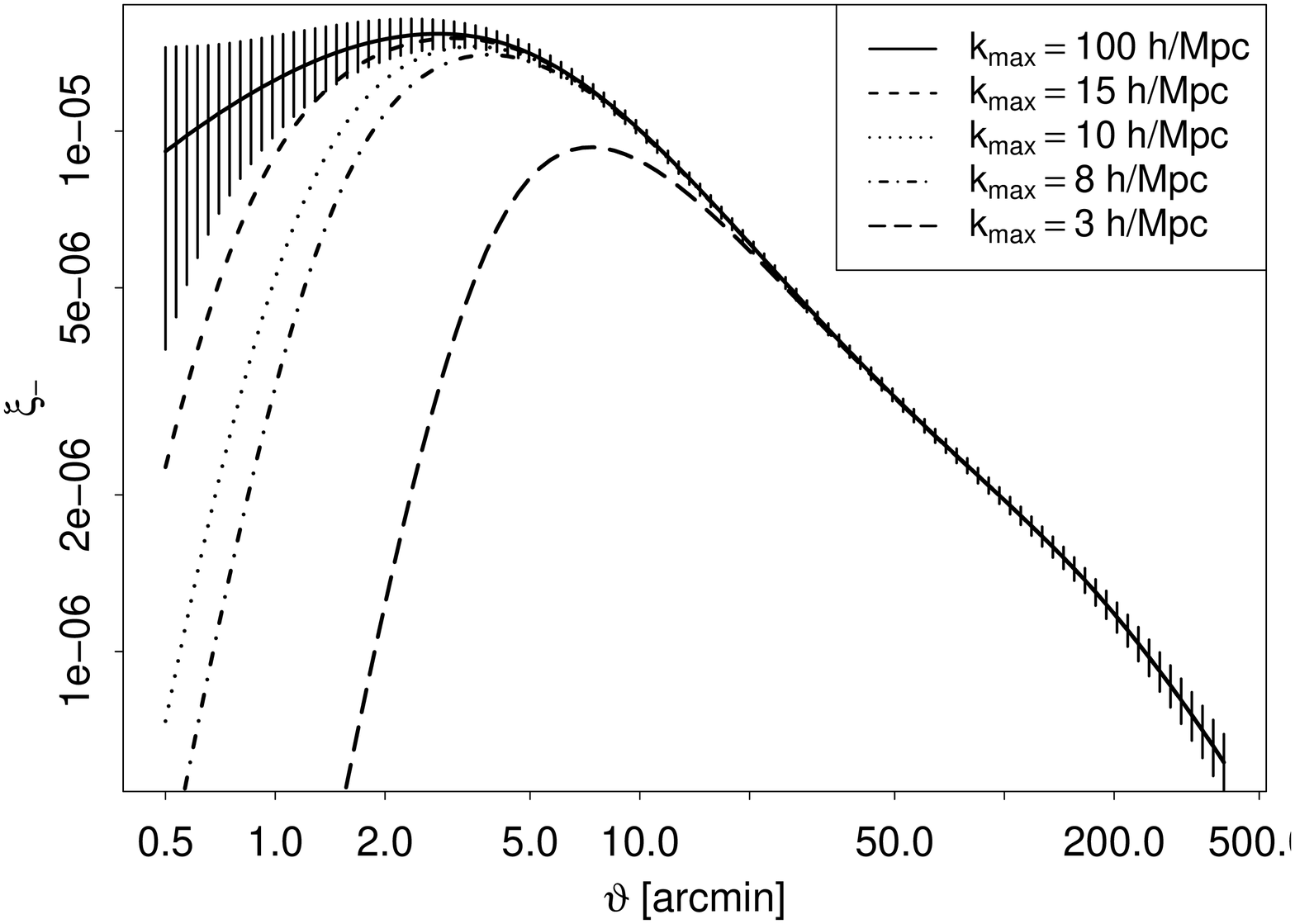}
\caption{\ti{Upper:} The relative error in the shear power spectrum $\pe(\kmax) / \pe(\kmax=100)$ for various cut-off scales $\kmax$ in the density power spectrum. The light shaded region indicates the statistical errors assumed for a DES-like survey, the dark shaded region corresponds to LSST-like error bars. \ti{Middle:} The 2PCF $\xip$ and \ti{Bottom:}$\xim$ (\ti{right}) for various $\kmax$ in $\pd$. Here we only show DES-like errors.}
\label{fi:Pkappa_xi_relative}
\end{figure}

For a future extension of the Coyote Universe project it is valuable to determine exactly over which scales $k$ and redshifts $z$ the density power spectrum must be known to the 1\% accuracy requirement defined in \cite{hut05}. Obviously the redshift range will depend on the redshift distribution of the considered survey; we outlined our choice in Sect. \ref{sec:pe} and argue that this is likely to be close to what is expected for DES. In Fig. \ref{fi:integrand_pe} we plot the ratio $\pe (z)/ \pe (z=4)$ as a function of $z$, where $\pe (z)$ is calculated as in Eq. \ref{eq:pdeltatopkappa} but with $w(z)$ as the upper limit. The shaded regions indicate statistical errors (intrinsic shape noise and cosmic variance) 
\be
\label{eq:delta_pe}
\Delta \pe = \sqrt{\frac{1}{\Delta \ell \,\ell \,(\ell+1) \, f_\mr{sky}}} \left(\pe+ \frac{\sigma_\epsilon^2}{2 \, n_\mr{gal}} \right)
\ee
for a DES-like survey (light shaded outer region, $f_\mr{sky}$=0.125, $n_\mr{gal}=12/\mr{arcmin^2}$), and a LSST/Euclid-like survey (dark shaded inner region, $f_\mr{sky}$=0.5, $n_\mr{gal}=40/\mr{arcmin^2}$). We choose $\sigma_\epsilon=0.3$ in both cases.\\ 
Here, the error bars are calculated at $\ell \sim 1000$ for the reason that they become minimal around this scale (see Fig. \ref{fi:Pkappa_xi_relative}). We find that for the considered redshift distribution, it is sufficient to know $\pd$ to $z \sim 1.5$ even for LSST like error bars. We note however that due to the increased depth of an LSST like survey its redshift distribution might be significantly different from what we assumed here.  \\
Next we impose a cut-off scale $\kmax$ in $\pd$, i.e. we set $\pd=0$ for $k>\kmax$, and study the impact of different choices of $\kmax$ on the shear power spectrum and the shear 2PCF. 
\begin{table}
\caption{The $\ell$ in $\pe$ where the error in the shear power spectrum $\Delta =| \pe(\kmax) / \pe(\kmax=100)-1|$ exceeds a certain percent level (see also Fig. \ref{fi:Pkappa_xi_relative}). The values of $k$ are given in h/Mpc.}
\label{tab:deltape}
\begin{tabular}{l l l l l}\hline
$\Delta $ & $\kmax=15$ & $\kmax=10$ & $\kmax=8$&$\kmax=3$ \\ \hline
1 $\% $ & $\ell=3593$ &$\ell=2096$ & $\ell=1557$& $\ell=377$  \\
5 $\%$ &$\ell=6604$ &$\ell=4152$ & $\ell=3152$& $\ell=823$ \\
10 $\%$ &$\ell=8589$ &$\ell=5483$ & $\ell=4213$& $\ell=1162$ \\
\end{tabular}
\end{table}
The top panel in Fig. \ref{fi:Pkappa_xi_relative} shows the relative error in the shear power spectrum $\Delta \pe=  |\pe(\kmax)/\pe(\kmax=100)|$ for various $\kmax$. 
We find that for DES-like error bars $\kmax=8 \mr{h/Mpc}$ is sufficient if one is only interested in $\pe (\ell)$ below $\ell=3960$, $\kmax=10 \mr{h/Mpc}$ extends this range to $\ell=5995$, and if $\pd$ can be modeled up to $\kmax=15 \mr{h/Mpc}$ even $\ell=15000$ is still within the DES error bars. This changes considerably for LSST, where knowing $\pd$ only up to $\kmax=8 \mr{h/Mpc}$ gives uncertainties in $\pe (\ell)$ above $\ell=1750$ that are equal/larger than the LSST errors. For $\kmax=10 \mr{h/Mpc}$ this threshold is shifted to $\ell>2308$, and for $\kmax=15 \mr{h/Mpc}$ to $\ell>4127$.\\
We perform a similar analysis for $\xip$ (Fig. \ref{fi:Pkappa_xi_relative}, middle panel) and $\xim$ (Fig. \ref{fi:Pkappa_xi_relative}, lower panel) which are calculated from $\pe$ via
\be
\label{eq:xipm}
\xi_\pm (\vt) = \int_0^\infty \frac{\d\ell\;\ell}{2\pi}\,{\mr J}_{0/4} (\ell\vt)
\left[P_{\rm E}(\ell)+P_{\rm B}(\ell) \right] \, .
\ee
The corresponding error bars for $\xi_\pm$ are derived from diagonal elements of the shear covariance $\Delta \xi=\sqrt{C^\xi_\mr{ii}}$, which can be calculated in terms of the power spectrum \citep{jse08}
\be
C^\xi_{\pm \pm}(\vt_i, \vt_j) = \frac{1}{4 \pi^2 f_\mr{sky}} \int_0^\infty  \d \ell \, \ell \, \mr J_{0/4}(\ell \vt_i) \, \mr J_{0/4}(\ell \vt_j) \left(\pe(\ell) + \frac{\sigma^2_\epsilon}{2 \, n_\mr{gal}} \right)^2 \,.
\ee  
The above expression only holds under the assumption that the four point function of the shear can be expressed in terms of power spectra which again only holds for Gaussian shear fields; including non-Gaussianity increases these error bars \citep{svh07, taj09}. More importantly, the 2PCF covariance has significant off-diagonal terms, hence its errors are strongly correlated which implies that the ratio of error bar to signal, as shown in Fig. \ref{fi:Pkappa_xi_relative}, does not fully resemble the actual signal-to-noise ratio of the 2PCF. For this one has to perform a full likelihood analysis taking the total covariance into account (see Sect. \ref{sec:like_analysis}). \\
Nevertheless our results in Fig. \ref{fi:Pkappa_xi_relative} show that $\xip$ is hardly affected by a cut-off even when choosing $\kmax=3 \mr{h/Mpc}$. In contrast the impact on $\xim$, especially on small scales, is much more severe. Even $\kmax=15$ h/Mpc is hardly within the DES error bars on scales $\leq 1'$.  The insensitivity of $\xip$ to the choice of $\kmax$ can be explained when looking Eq. \ref{eq:xipm} and the filter function $\mr J_0 (\ell \vt)$. We see that the main contribution to the integral in Eq. \ref{eq:xipm} comes from small $\ell$ (where $\pe$ is not affected by a cut-off in $\pd$), and that this tendency increases for larger arguments of $\mr J_0$, i.e. larger $\vt$. In contrast, for $\xim$ we see that $\mr J_4 (\ell \vt)$ strongly suppresses these small $\ell$ and collects power mainly from larger scales, which is the reason why it is more biased when using low $\kmax$.

\section{COSEBIs and ring statistics}
\label{sec:cosebis}
We extend the analysis to the most recently developed EB-decomposition methods, namely the ring statistics and the COSEBIs. Both can be expressed as integrals over the 2PCF, i.e.
\be
\label{eq:ring}
\rrEB (\vt)=\int_{\vt_\mr{min}}^\vt  \frac{\mr d \vt'}{2\,\vt'} \left[ \xip(\vt') \, Z_+(\vt',\vt) \pm \xim(\vt') \, Z_-(\vt',\vt)\right], \\
 \ee 
for the ring statistics, and
\be
\label{eq:cosebis}
E_n/B_n= \int_\tmin^\tmax \frac{\d\vt}{2}\;\vt\,\left[\tplog(\vt) \, \xi_+(\vt) \pm \tmlog(\vt) \, \xi_-(\vt) \right],
\ee
in case of the COSEBIs. The filter functions $Z_\pm$ are derived in \cite{sck07}, we use the modified version described in \cite{esk10}. The derivation of the COSEBIs' filter functions $T_{\pm n}$ is outlined in detail in SEK10. The superscript $^\mr{log}$ indicates that the roots of the corresponding filter function are distributed logarithmically in $\vt$. We note that a similar relation exists for linear $T$-functions, which can be expressed very conveniently in terms of Legendre polynomials. However $\tpmlog$ has the advantage to compress the information contained in the 2PCF into significantly fewer data points hence we consider only $\tpmlog$ in this paper.\\
Another way to calculate the COSEBIs EB-mode employs the shear E-mode power spectrum, or a B-mode power spectrum, respectively 
\be
\label{eq:Epowdef}
E_n/B_n=\int_0^\infty \frac{\d\ell\;\ell}{2\pi} P_{\rm E/B}(\ell)\,W^\mr{log}_n(\ell) \;, 
\ee
where the filter functions $W^\mr{log}_n$ can be calculated from the $\tpmlog$ as
\bea
\label{WnFilter}
W^{\mr{log}}_n(\ell)&=&\int_{\tmin}^{\tmax}\d\vt\;\vt \, T^{\mr{log}}_{+ n}(\vt) \, \mr J_0(\ell \vt) \, \\
&=&\int_{\tmin}^{\tmax}\d\vt\;\vt \, T^{\mr{log}}_{- n} (\vt) \, \mr J_4(\ell \vt) \;.
\eea
A similar relation exists for the ring statistics but using $Z_\pm$ instead of $T^{\mr{log}}_\pm$.\\
Similarly to Eqs. (\ref{eq:cosebis}, \ref{eq:Epowdef}) the covariance of the COSEBIs' (and ring statistics') E-mode can be expressed through the covariance of the $\xipm$, and the covariance of $\pe$ 
\bea
\label{eq:CEpower}
C^{\mr E}_{mn}&=&\frac{1}{4 \pi^2 f_\mr{sky}} \int_0^\infty \d\ell\;\ell\,W^{\mr{log}}_m(\ell) W^{\mr{log}}_n(\ell)
\left( P_\mr E(\ell)+\frac{\sigma_\eps^2}{2\, n_\mr{gal}} \right)^2 \,, \\
\label{eq:CExi}
&=&\frac{1}{4}\int_{\tmin}^{\tmax}\d\vt\;\vt
\int_{\tmin}^{\tmax}\d\vt'\;\vt' \nonumber \\
&\times&\sum_{\mu,\nu=\{+,-\} }
T^{\mr{log}}_{\mu m}(\vt)\, T^{\mr{log}}_{\nu n}(\vt')\, C_{\mu\nu}(\vt,\vt')\;.
\eea
\begin{figure*}
\includegraphics[width=8cm]{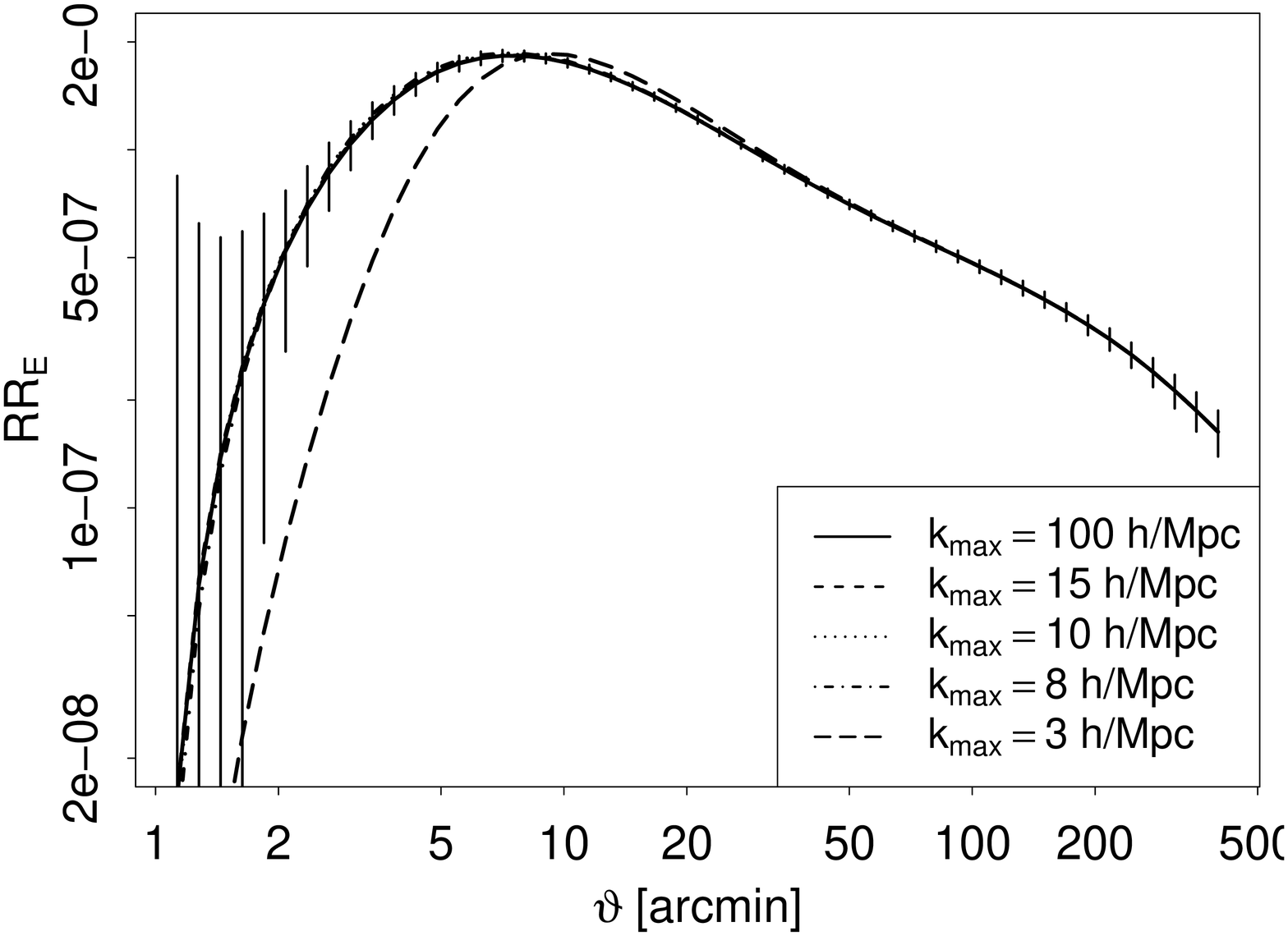}
\includegraphics[width=8cm]{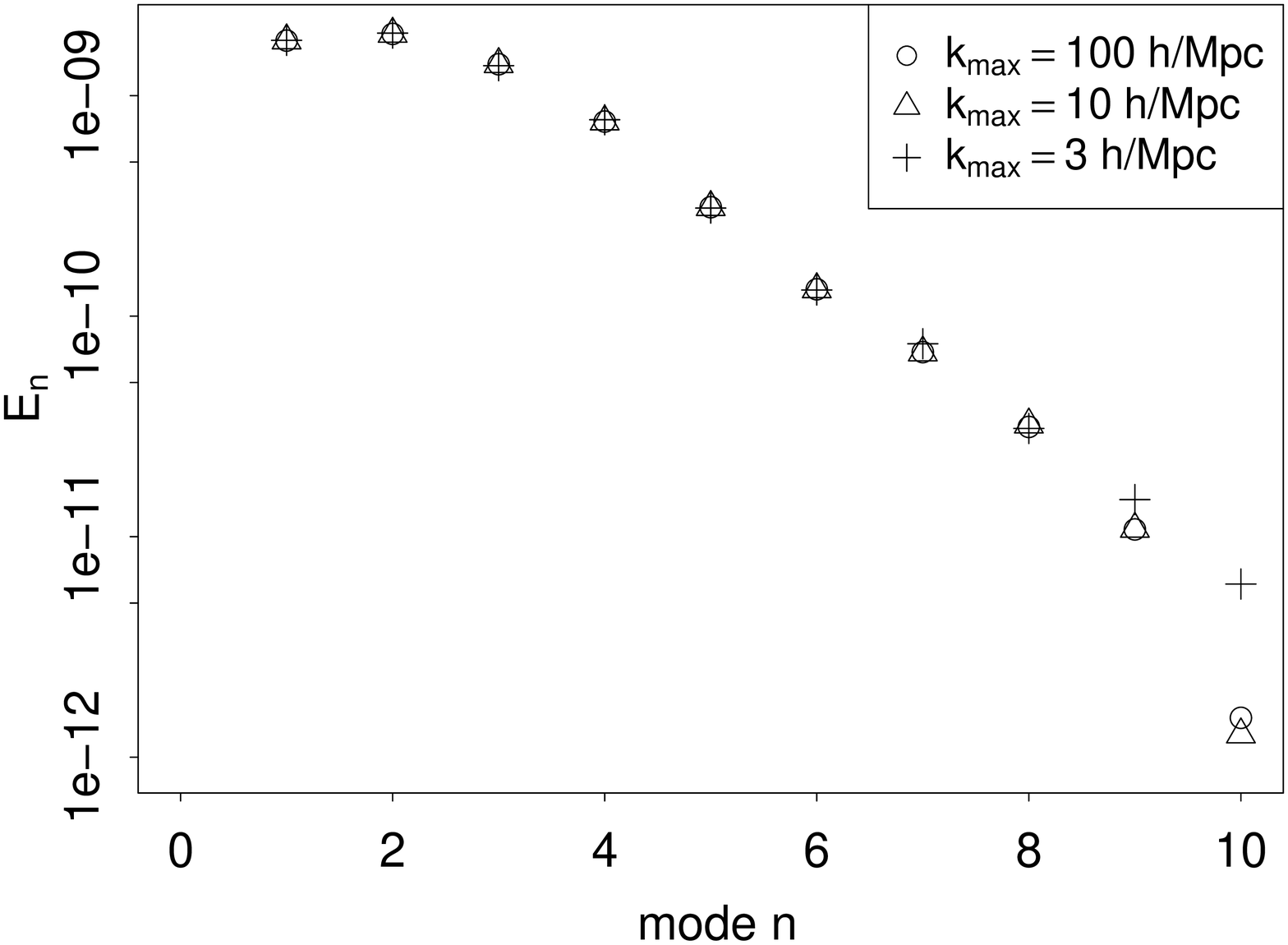}
\caption{The ring statistics (\textit{left}) with DES-errors and the COSEBIs (\textit{right}) for various $\kmax$ in $\pd$. Note that we refrain from plotting errors for the COSEBIs as they are highly correlated (see text).}
\label{fi:EB_cut}
\end{figure*}
In Fig. \ref{fi:EB_cut} we perform a similar analysis as in Fig. \ref{fi:Pkappa_xi_relative} but for the ring statistics (\ti{left panel}) and for the COSEBIs (\ti{right panel}). The error bars for the ring statistics are calculated as the square root of the diagonal elements of the ring statistics covariance. In contrast to the 2PCF this gives a fair estimate of the signal-to noise ratio, as the ring statistics covariance is almost diagonal \citep{esk10}. We find only a small bias depending on $\kmax$; only for $\kmax=3$ h/Mpc there is a clear deviation, all other estimates of the ring statistics signal are well within the DES error bars. This behavior can be explained again when looking at the ring statistics as a filtered version of the shear power spectrum, and the corresponding filter function $W (\ell)$ \citep[][Fig. 3]{sck07}. The filter function suppresses the large $\ell$ information, where the shear power spectrum is most affected by a low $\kmax$.\\
We perform a similar analysis for the COSEBIs, however refrain from plotting the error bars as they are heavily correlated, at least for the logarithmic case. We postpone an error analysis for the COSEBIs to the next section, where we carry out a complete likelihood analysis in a 5-dimensional cosmological parameter space, that takes the full COSEBIs covariance into account. Nevertheless, one can see that the COSEBIs can be calculated accurately up to the 9th mode for  $\kmax=10$ h/Mpc. Choosing $\kmax=3$ h/Mpc results in a notable difference from the 7th mode on. It is the major purpose of the next section to check how many modes are needed to capture the cosmological information of second-order shear measures in a 5-dimensional parameter space.

\section{Likelihood analysis}
\label{sec:like_analysis}
In SEK10 we analyze the performance of the COSEBIs relative to the theoretical 2PCF for the $\om$-$\sig$ parameter space. Here, we extend this analysis in several ways: First, we consider a parameter space consisting of 5 parameters ($\omega_m, \omega_b, \sig, \ns, \w$) to quantify how many modes $E_n^\mr{log}$ are needed to capture the bulk of cosmological information in this increased parameter space. Note that we use $\omega_m=\om h^2$ ($\omega_b=\omb h^2$) instead of $\om$ ($\omb$), for the reason explained in Sect. \ref{sec:predictions}. Second, we compare the information content of the COSEBIs when calculated from a 2PCF that is measured on only small scales ($\vt \in [1';100]$) and only large scales ($\vt \in [120' ; 400']$). This analysis is motivated by the findings in SEK10 (see Fig. 11), namely that for $\tmin=1'$ the information content of the COSEBIs does not increase significantly when increasing $\tmax$ beyond $200'$ .\\ 
The data vectors of the COSEBIs are calculated from the 2PCF via Eq. (\ref{eq:cosebis}), and we cross-check the results using Eq. (\ref{eq:Epowdef}). Similarly we calculate the covariance using both Eqs. (\ref{eq:CExi}) and (\ref{eq:CEpower}). For covariances this consistency check is more important as numerical difficulties arise more likely in the covariance calculation (see Appendix of SEK10 for details). The survey parameters we assume in the calculation of all covariances in this paper correspond to those of the Dark Energy Survey (see Sect. \ref{sec:predictions}).
\subsection{Likelihood formalism}
\label{sec:formalism}
\begin{figure}
\includegraphics[width=9cm]{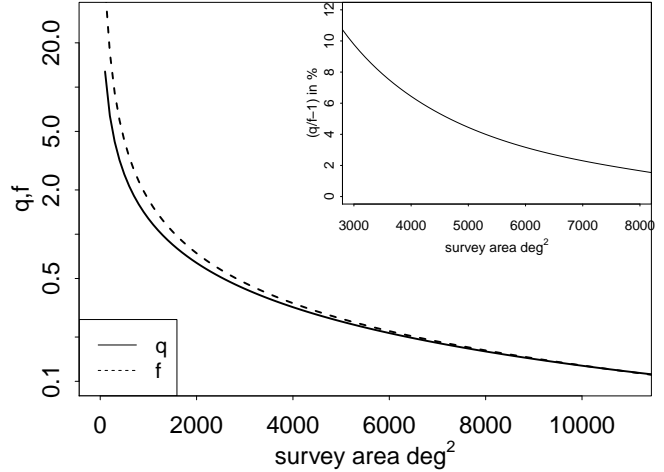}
  \caption{The dependence of $q$ and $f$ on the survey size for the $\om$-$\sig$ parameter space only. This illustrates the difference of a fisher matrix based figure-of-merit and a figure-of-merit that is based on a full likelihood analysis.}
         \label{fig:qf_compare}
\end{figure}
In this paper we quantify the information content of a measure using the figure-of-merit 
\be
\label{q_def}
q = \sqrt{|\matQ|} \quad \mr{with}\quad  Q_{ij} = \int \mr d^2 \vpi \, p(\vpi_{ij}| \,\vecd) \; (\pi_i-\pi_i^{\mr f})(\pi_j -\pi_j^{\mr f}) \,.
\ee
In our case $Q$ is a 5 $\times$ 5 matrix, which can be interpreted as the expectation value of the parameter covariance matrix. The parameter vector reads $\vpi=(\omega_m,\omega_b,n_s,\sig,w_0)$, and $\pi_i^{\mr f}$ and $\pi_j^{\mr f}$ denote the fiducial parameter values. Note that the posterior likelihood $p(\vpi_{ij}|\vecd)$ is calculated by marginalizing over the likelihood of the three parameters $\neq \pi_i, \pi_j$.\\
The figure-of-merit $q$ corresponds to the more common Fisher matrix based figure of merit $f=1/|\sqrt{\mathbf F}|$ \citep[see][ SEK10 for the exact definition]{tth97} if the likelihood in parameter space is Gaussian. The Fisher matrix $\mathbf F$ can be interpreted as the expectation value of the \textit{inverse} parameter covariance evaluated at the maximum likelihood estimate parameter set, which in our ansatz corresponds to the fiducial parameters. Mathematically we can express this equivalence as
\be
\label{eq:qf_equi}
f=\frac{1}{\sqrt{|\mathbf F|}}=\sqrt{|\matC_{\vpi}|}=\sqrt{|\matQ|}=q \,.
\ee
However, the assumption of a Gaussian likelihood in parameter space is only justified close to the maximum of the likelihood function, i.e. close to the fiducial parameters. If the likelihood function does not fall off quickly enough, $f$ is a rather bad approximation and can give significantly different values compared to $q$ (see SEK10). Obviously the Gaussian likelihood assumption significantly becomes more valid the larger the survey volume and we examine this behavior quantitatively in Fig. \ref{fig:qf_compare} for the two-dimensional parameter space $\om$-$\sig$.  We find that $q$ and $f$ deviate significantly for small survey volumes, whereas this difference drops below $1 \%$ for surveys $>$ 9300 $\mr{deg}^2$. Here, we consider a survey size corresponding to the Dark Energy Survey, i.e. $5000 \mr{deg}^2$ where the deviation is still $>$7 \%; therefore we refrain from using $f$ in our analysis and consider $q$ only. \\
In the calculation of the posterior likelihood we assume that the data points in $\vecd$ (which in our case are either COSEBIs or 2PCFs) are following a Gaussian error distribution 
\be
\label{likelihood}
\mathcal L=\frac{1}{(2\pi)^{N/2}\,\sqrt{|\matC|}} \, \mr{exp} \left[ - \frac{1}{2}\, (\vecd^f - \vecd (\vpi))^t \, \matC^{-1} \, (\vecd^f - \vecd (\vpi)) \right] \;, 
\ee
and that the covariance $\matC$ is constant in parameter space. These assumptions are not fully justified (see \cite{hss09,ssd10} for information on the first and \cite{esh09} for information on the second assumption) and they need to be examined carefully when inferring cosmological constraints from data. For our purpose these assumptions should not influence the results qualitatively as we are comparing the information content of 2PCF and COSEBIs relative to each other.\\
\begin{figure*}
 \includegraphics[width=8cm]{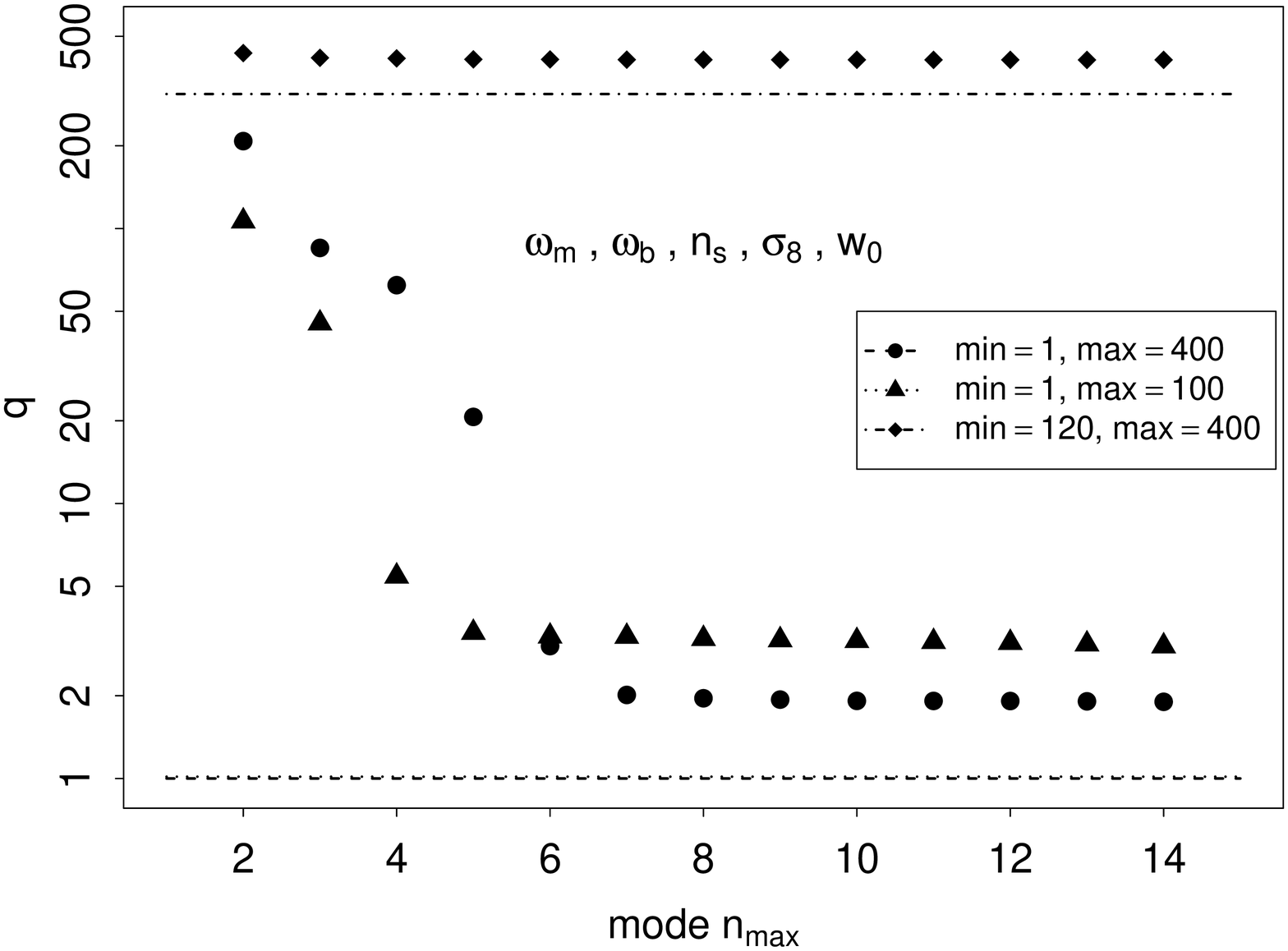}
 \includegraphics[width=8cm]{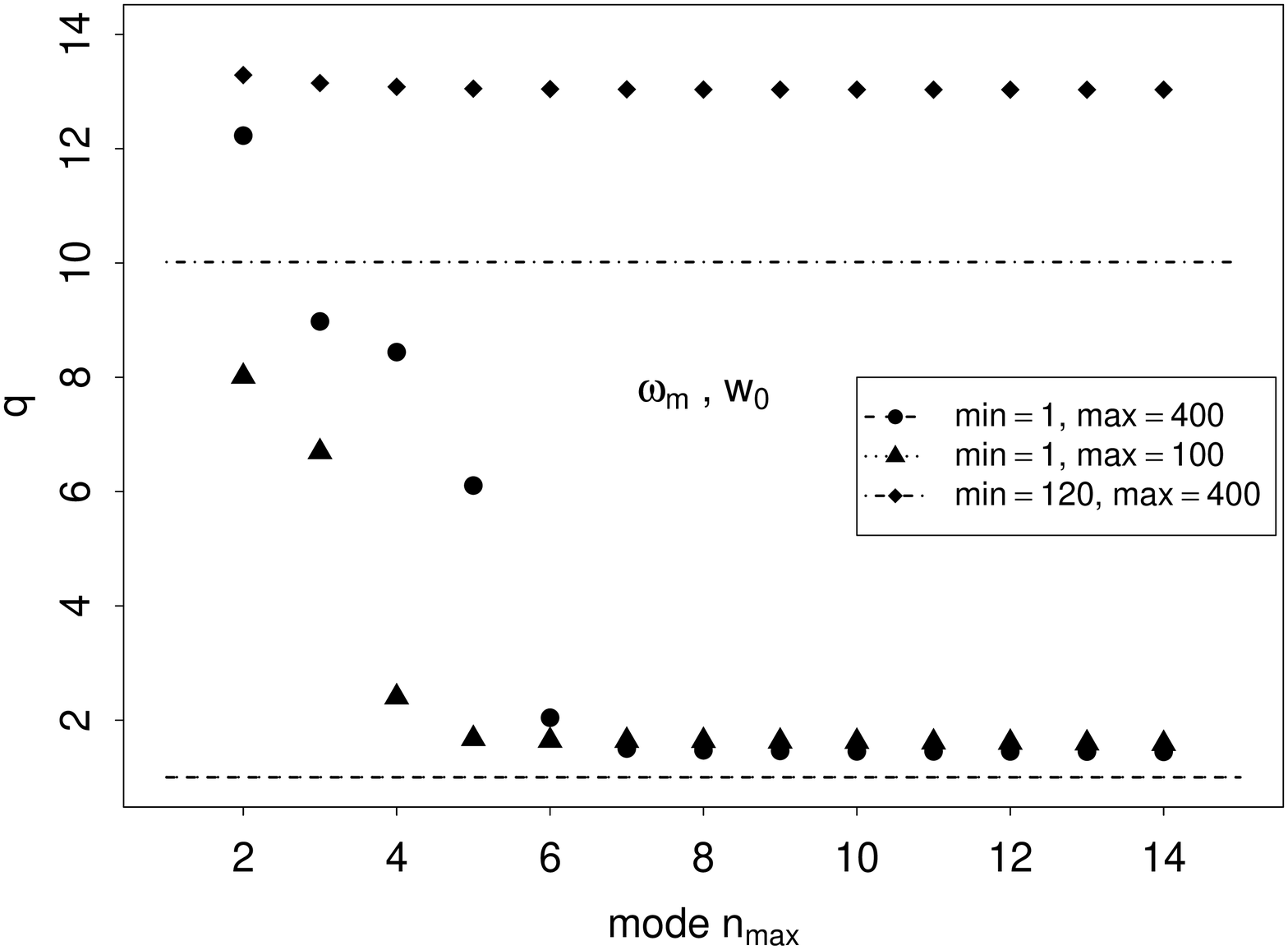}
 \includegraphics[width=8cm]{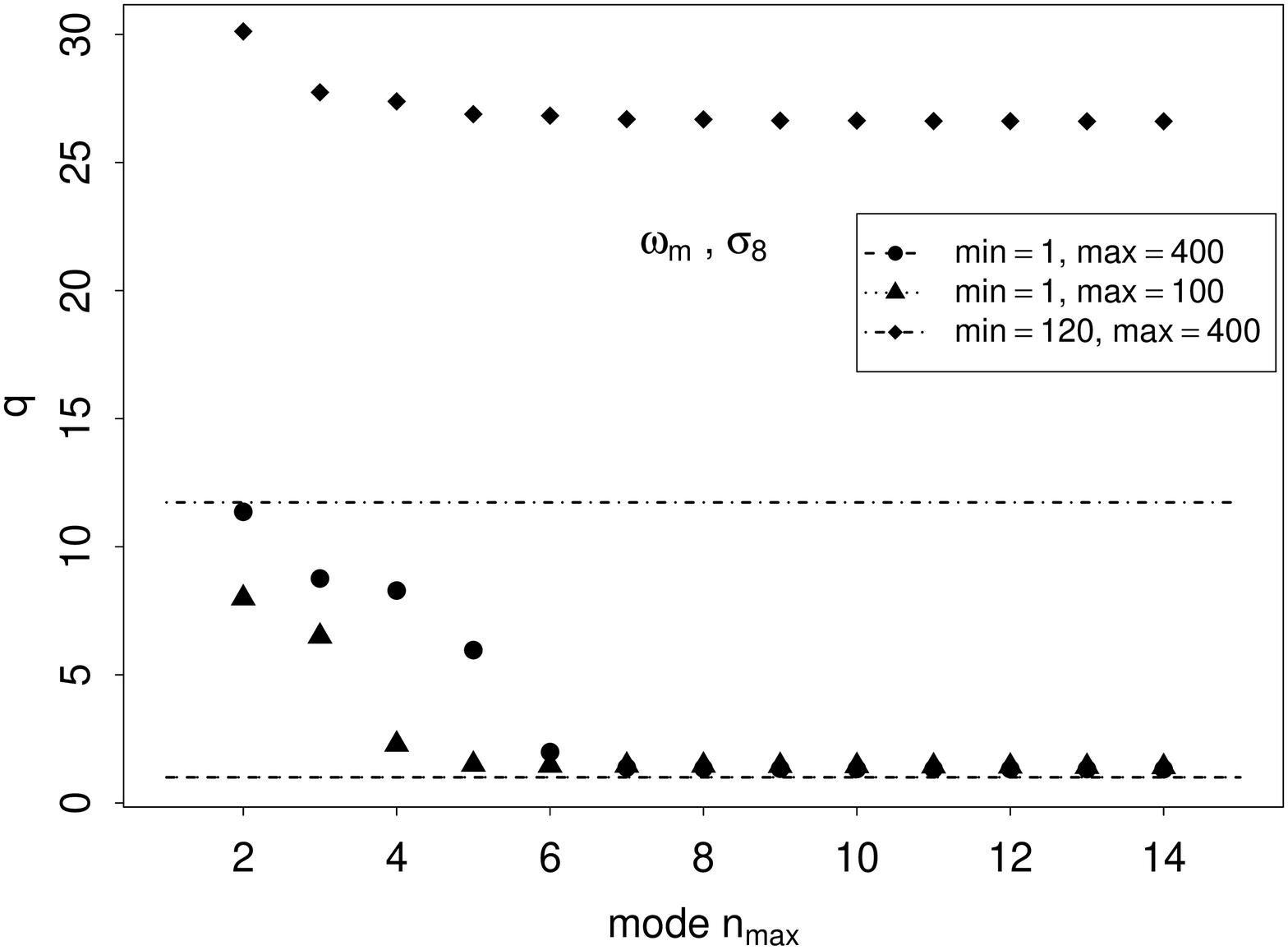}
 \includegraphics[width=8cm]{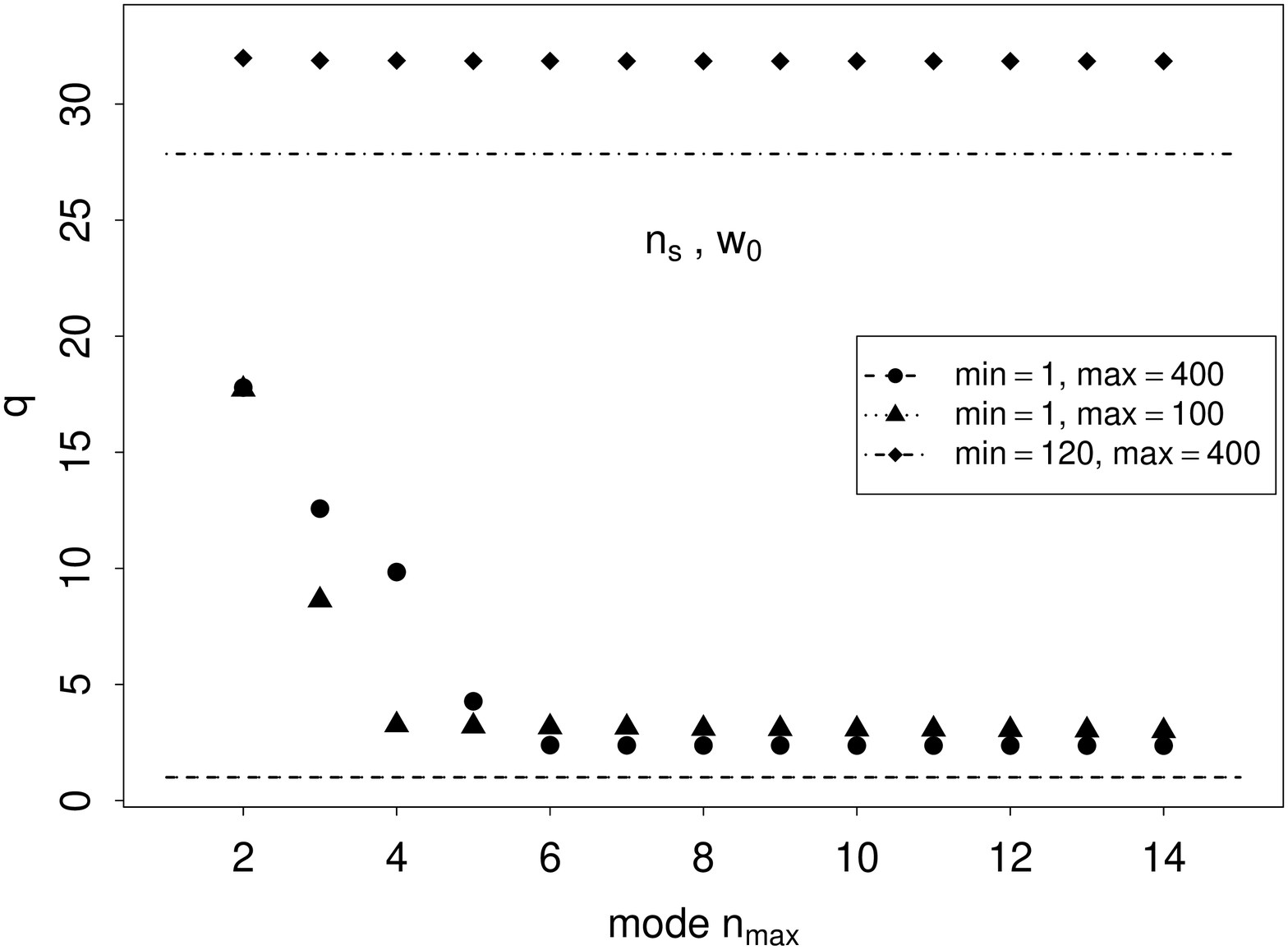}
\caption{The information content of the 2PCF and the COSEBIs for the 5-dimensional parameter space (\ti{top left}) and for three two-dimensional parameter spaces quantified by the measure $q$. Note that smaller $q$corresponds to larger information content (smaller likelihood contours). The lines correspond to the 2PCFs' $q$ for the 3 considered $\vt$-intervals and pose the upper limit of the second order information content. The values of $q$ for the COSEBIs are plotted as a function of number of modes $n$, that are included in the likelihood analysis.}
\label{fi:q_marg}
\end{figure*}

\subsection{Results}
\label{sec:results}
In Sect. \ref{sec:formalism} we describe our figure of merit as the determinant of the parameter covariance matrix $\matQ$. In addition, to the full parameter covariance we calculate the determinants of several submatrices of $\matQ$; these $q$ can be illustrated as two-dimensional likelihood contours where we marginalize over the other parameters.\\
In Fig. \ref{fi:q_marg}, we show the $q$ of the COSEBIs compared to the pure E-mode 2PCF for the 5-dimensional parameter space (\ti{top left}) and when considering the subsets. $\omega_m$ vs. $w_0$ (\ti{top right}), $\omega_m$ vs. $\sig$ (\ti{bottom left}), $n_s$ vs. $w_0$ (\ti{bottom right}). We performed the same analysis for the other 6 parameter subsets, but only show this selection due to the fact that the results look very similar. Note that scale of $q$ is logarithmic in the top left panel whereas it is linear in all other cases. In addition, we scale all $q$ such that $q=1$ for the $[1';400']$-2PCF(dashed line).   \\
In each panel we consider three intervals in angular scale $[1';400']$, $[1';100']$, and $[120';400']$, and calculate the value of $q$ for the E-mode 2PCF as the upper limit on  cosmological information from second-order shear measures. Note that the line representing the $[1';100']$-2PCF overlaps with the line of the $[1';400']$-2PCF. We then calculate the $q$ of the COSEBIs as a function of modes that are included in the likelihood analysis, and thereby quantify their cumulative cosmological information as a function of modes.\\
Considering the $[1';400']$-interval we find that even in a 5-dimensional parameter space the number of COSEBI-modes that is required to capture the cosmological information is quite small, $\sim$7-8 modes (\ti{top left}). This guarantees a manageable covariance matrix and a stable calculation of its inverse. In addition, we see that knowing the density power spectrum to $\kmax=10$ h/Mpc is sufficient to avoid uncertainties in the data analysis coming from theoretical predictions. The difference of the 8-mode COSEBIs to the shear 2PCF is very small, hence the loss in information is a negligible inconvenience compared to the uncertainties that are avoided when using the COSEBIs instead of the 2PCF.\\
Slightly surprising is the difference in information when comparing the $[1';100']$- and the $[120';400']$-interval. The $[1';100']$-COSEBIs' information content saturates at $\sim$5 modes for 5 cosmological parameters, and the difference to the $[1';400']$-case is rather small. Comparing the $q$ of the $[1';100']$- and $[1';400']$-2PCF we can hardly see a difference, which indicates that the bulk of cosmological information is indeed contained on small scales. The $[120';400']$-COSEBIs saturate already at 2-3 modes, and the difference to the other saturation limits is large. Also the difference between COSEBIs and 2PCF is much more significant in this case compared to the other two. \\
To explain this behavior one needs to examine the variation of the signal-to-noise ratio of the 2PCF with respect to cosmology. Knowing that the bulk of the cosmological information is contained in $\xip$, and assuming that the square root of the diagonal of the covariance serves as a rough estimate for the noise-level, we see from Fig. \ref{fi:Pkappa_xi_relative} that the signal-to-noise is relatively low on scales $>$100' compared to the scales $<$100'. Plotting the variation of the 2PCF signal in the 5-dimensional cosmological parameter space we find the relative deviation to the signal of the fiducial cosmology is roughly constant over all scales. The lower noise level on small scales then implies that these scales contribute more information on cosmology.\\  
There are only slight variations to this behavior when looking at the other panels of Fig. \ref{fi:q_marg}. As expected the information content of the COSEBIs saturates earlier compared to the 5-dimensional parameter case, however the fact that almost all the information is present on scales $<$100' remains. 

\section{Conclusions}
\label{sec:conclusions}
Accurate predictions of cosmic shear measures play an important role in the analysis of future weak lensing data sets. In the first part of this paper we present our new weak lensing predictions pipeline that is based on the Coyote Universe emulator and extends the emulator using the Halofit code. A shear power spectrum derived from this new pipeline differs from a shear power spectrum calculated from Halofit only by 6-11 \%, depending on the Fourier mode $\ell$.\\
We consider this as a first step in developing a weak lensing predictions pipeline that meets the requirements for DES and later for LSST, Euclid, and WFIRST. For this we examine to which $k$ in h/Mpc and $z$, the density power spectrum must be modeled accurately by numerical simulations in order to meet those requirements. We find that $\kmax=8$ $\mr{h/Mpc}$ causes a bias in the shear power spectrum at $\ell=4000$ that is within the statistical errors (intrinsic shape-noise and cosmic variance) of a DES-like survey, whereas for LSST already $\kmax=15$ $\mr{h/Mpc}$ is needed. \\
A future pipeline for weak lensing predictions that models the density power spectrum to such large $k$, the shear power spectrum to such high $\ell$, respectively, will have to take baryons into account \citep{jzl06,rzk08}. For example, \cite{jzl06} find that the shear power spectrum on scales of $\ell \in [1000,10000]$ are affected by 1-10\%, depending on the level of complexity in the treatment of baryons.\\
In addition to the treatment of baryons additional corrections need to taken into account. For example, shape distortions probe the reduced shear $g=\gamma/(1-\kappa)$ instead of the shear ($\gamma$) itself \citep[e.g.,][]{sha09}, and when calculating the reduced shear from the measured mean ellipticity higher order terms in $g$ need to be taken into account. Corrections to the Born approximation are necessary to account for multiple deflections of light rays \citep[e.g.,][]{hhw09}, and corrections to biases that occur because de(magnification) of galaxies due to lensing correlates with selection criteria of the considered galaxy sample \citep[e.g.,][]{srd09}. These corrections are calculated in \cite{krh10}; the authors find that reduced shear and magnification bias are important already for DES-like surveys, whereas the other effects will only become important for weak lensing data from LSST.\\
In the second part of this paper we extend earlier studies of the COSEBIs, the most recently developed EB-mode decomposing second-order cosmic shear measure. In particular, we are interested in their performance in a high-dimensional cosmological parameter space. Compared to other second-order shear statistics the COSEBIs can be calculated from a shear 2PCF measured on a finite interval $[\tmin;\tmax]$. The 2PCF again is independent of any masking effects or survey geometry and thereby avoids several difficulties present in other second-order shear measurement methods, e.g. the shear power spectrum. \\ 
The COSEBIs can be imagined as a filtered version of the 2PCF (or the shear power spectrum); instead of having angular scale $\vt$ or Fourier modes $\ell$ as an argument, they are a function of the order of the polynomial which is used as their filter function.  Furthermore, they are designed to condense the second-order cosmic shear information into a (small) number of discrete modes. In SEK10 two different filter functions are examined: Filter functions that sample the 2PCF linearly can be expressed very conveniently in terms of Legendre polynomials, however the number of modes that is needed to capture all the cosmological information is rather large in this case. The logarithmic filter functions are much more efficient; SEK10 find that $\sim$ 5 modes are sufficient to capture the cosmological information in the two-dimensional parameter space $\om$ vs. $\sig$.\\   
Here, we extend their analysis and examine the performance of the COSEBIs in a 5-dimensional parameter space. We find that the bulk of the cosmological information is contained on angular scale $<$100', whereas scales above this threshold only contribute little to constraints on cosmology. Furthermore, we find that $\sim$8 modes of the COSEBIs are sufficient to capture all the EB-mode decomposable cosmological information. This number is still relatively small, in particular as we show that the COSEBIs can be calculated accurately up to the 9th mode when knowing $\pd$ ``only'' up to $\kmax=10$ h/Mpc. From these results we conclude that the COSEBIs are fairly robust against theoretical uncertainties in modeling the density power spectrum, and represent a good choice as a second-order weak lensing statistics.

\section*{Acknowledgments}

I thank Katrin Heitmann, Salman Habib, and Martin White for the help with the Coyote Universe emulator and the useful discussions. I also want to thank Peter Schneider and Katrin Heitmann for reading the manuscript and giving helpful comments.


\label{lastpage}
\end{document}